\documentclass[twocolumn,showpacs,preprintnumbers,amsmath,amssymb,10pt,a4paper]{revtex4}
\pdfoutput=1
\usepackage{geometry}
\usepackage{graphicx}% Include figure files
\usepackage{dcolumn}% Align table columns on decimal point
\newcommand{\be}{\begin{equation}}
\newcommand{\ee}{\end{equation}}
\newcommand{\bd}{\begin{displaymath}}
\newcommand{\ed}{\end{displaymath}}
\newcommand{\BE}{\begin{eqnarray}}
\newcommand{\EE}{\end{eqnarray}}

\newcommand{\avg}[1]{\left\langle{#1}\right\rangle}

\begin{document}

\preprint{}
\title{Effects of communication and utility-based decision making in a simple model of evacuation}
% Force line breaks with \\

\author{Michalis Smyrnakis}
\email{michalis.smyrnakis@manchester.ac.uk}

\author{Tobias Galla}
\email{tobias.galla@manchester.ac.uk}

\affiliation{Theoretical Physics, School of Physics and Astronomy, The University of Manchester, Manchester M13 9PL, United Kingdom}

\date{\today}% It is always \today, today,

\begin{abstract}
We present a simple cellular automaton based model of decision making during evacuation.  Evacuees have to choose between two different exit routes, resulting in a strategic decision making problem. Agents take their decisions based on utility functions, these can be revised as the evacuation proceeds, leading to complex interaction between individuals and to jamming transitions. The model also includes the possibility to communicate and exchange information with distant agents, information received may affect the decision of agents. We show that under a wider range of evacuation scenarios performance of the model system as a whole is optimal at an intermediate fraction of evacuees with access to communication. 
\end{abstract}

\maketitle
%%%%%%%%%%%%%%%%%%%%%%%%%%%%%%%%%%%%%%%%%%%%%%%%%%%%%%%%%%%%%%%%%%%%%%%%%%%%%%%%
%%%%%%%%%%%%%%%%%%%%%%%%%%%%%%%%%%%%%%%%%%%%%%%%%%%%%%%%%%%%%%%%%%%%%%%%%%%%%%%%
\section{Introduction}
The theory of complex systems originated in physics and chemistry several decades ago, and has traditionally focused on topics such as self-organisation \cite{haken,bak,kauffman}, pattern formation \cite{cross}, and other emergent properties of multi-particle systems interacting at the micro-level. Quickly these ideas have found applications in other areas, including for example the modelling of the brain \cite{hopfield}, that of virus dynamics and the immune system \cite{may}, social dynamics and decision making \cite{fortunato} and economics \cite{stanley}. One of these applications is the modelling of transport processes. This can refer to the transport in biological systems, for example intracellular transport of molecules, signal transduction or models of DNA transcription and translation. Complex systems approaches to modelling biological transport have also been applied to the dynamics of swarms or traffic in colonies of social insects, e.g. ants. In the context of humans models of transport are relevant for the understanding of vehicular traffic, pedestrian motion and they have been applied to model evacuation procedures. A recent textbook overview with comprehensive references can be found in \cite{schadschneider}, see also \cite{helbingreview,helbingreview2,SKK09}.

Transport models typically come in two different types. The first type consists of basic simple models, aiming to capture the basic phenomenological effects without too much ambition of describing all details of the underlying real-world processes. Frequently such models are formulated in terms of cellular automata (CA) \cite{wolfram}, space is discretized into different cells, who can each be in one of a small number of states at each time (e.g. `occupied' or `vacant'). This constitutes a significant degree of coarse graining. Simple rules are then devised to propagate particles in space, and hence modifying the states of cells.  Examples of such systems include the celebrated Nagel-Schreckenberg model of vehicular traffic \cite{nagel}, simple models of biological transport based on exclusion processes \cite{macdo1,macdo2}, and various CA-based models of pedestrian motion. Simple CA-models include in particular the models proposed by Fukui and Ishibashi \cite{fukui1,fukui2}, the one by Blue and Adler \cite{blue}, an overview can be found in \cite{schadschneider}. More sophisticated CA-models take into account interactions between evacuees beyond a simple exclusion rule, and augment these by so-called `floor fields', akin to pheromone traces in chemotaxis. Floor fields have a static component, representing properties of the underlying infrastructure, and a dynamic component, reflecting the effects of the changing positions of evacuees. A good review can again be found in \cite{schadschneider}.

The second class of transport models consists of more detailed descriptions of the underlying real-world dynamics. For example there exist evacuation models of various structures, cities and areas \cite{lsm}, describing the evacuation processes in some detail, including features such as populations with age-structures, different mobilities, combinations of different means of transport and so on. On the technical level the social force model first proposed by Helbing and co-workers \cite{HFV00} might be seen as more fundamental than the above CA-based models. Such models directly start from the particles (individuals) to be propagated in space and formulate the forces acting between them. Propagation then occurs via Newton's law. Typically two types of forces are taken into account: physical forces between the individuals and/or the surrounding structures (e.g. a wall) and social forces, such as crowding effects. This stresses the importance of combining two distinct aspects of evacuation dynamics, physical processes (i.e. motion of evacuees) and psychological processes such as crowd dynamics, decision making and communication. In \cite{schadschneider} these two aspects are referred to as the `tactical level', denoting the short-term decisions agents make and the `operational level', describing the actual motion of evacuees in space. The focus of our work is mostly the tactical level, we aim to investigate how decision making and in particular communication between agents affects the outcome of evacuation in simple models of pedestrian motion. Our aim is not to improve models on the operational level, instead we choose a basic cellular automaton, to be described below.

In the context of our model decision making mostly refers to choosing one of several exit routes. Several approaches have been proposed in the existing literature as methods for the evacuees to select an available exit. These approaches use for example the general structure of the area being evacuated via floor fields \cite{ff1,ff2},  or the current configuration of agents \cite{rule3} as an input for decision making. Decision making rules that have been used include neural networks \cite{neural1, neural2}, best response dynamics and logit based models \cite{game, logit} and rule-based methods that focus on different attributes of the floor field \cite{rule1, rule2,rule3,rule4,rule6,ff1}.  Existing models of communication typically address the exchange of information about the area to be evacuated  either between evacuees \cite{commun1} or using a leader-follower approach where the leader informs the followers as in \cite{commun2}. 

In the present paper we discuss a model with a different communication approach focusing on the effects of direct communication between the agents on their decision making and choosing between different exit routes, and the effect on evacuation efficiency. In these communication events the agents exchange information about the current flow velocities they experience towards their respective destinations. We use a simple CA-based model to describe the motion of pedestrians in an evacuation scenario with several exit routes. Evacuees then have to make a decision which exit to choose. This naturally leads to a competitive situation, not too dissimilar to those in competive minority games \cite{mg}. Even if one exit is objectively more desirable (e.g. wider) not all agents can evacuate through this exit, and overall global performance might improve if some agents evacuate via the less desirable route. The decision making in our model is based on utility functions each individual assigns to each of the exits. These utilities are based on the distance to the exit, but also on the speed with which the agent moves. Hence utilities may change with time, and decisions may be revised. Communication between agents may alter the perceived utilities. An agent may for example receive information on a distant exit route (e.g. via mobile telephones), and as a consequence he may change his current choice of exit route. As one main result of our work we demonstrate that the overall evacuation performance can attain optimal values at intermediate degrees of communication. I.e. evacuation is comparatively slow if either all or no agents have access to communication, but with better performance at an intermediate fraction of evacuees who can communicate. The remainder of the paper is organised as follows: In Sec. \ref{sec:model} we introduce the model dynamics. Sec. \ref{sec:res} then contains our main simulation results before we present a summary and an outlook in Sec. \ref{sec:concl}

\section{Model definitions}\label{sec:model}
\subsection{General setup}
\begin{figure}[t!!]
\vspace{-1em}
\centerline{\includegraphics[scale=0.55]{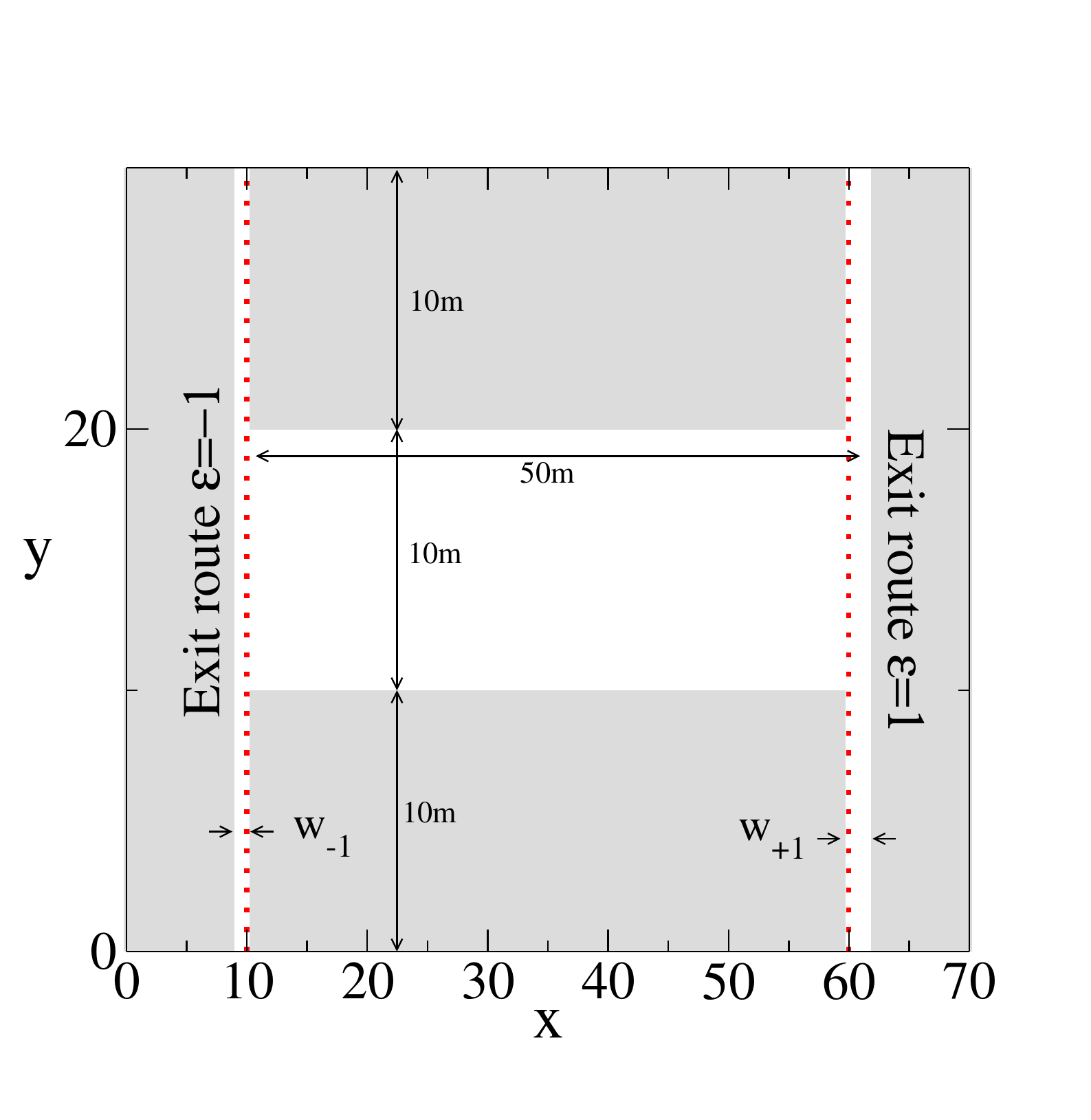}}
\caption{Spatial structure underlying our simulations. One central area of size $50$m$\times 10$m is initially populated with agents, who then evacuate through one of two exit routes, of widths $w_{-1}$ and $w_{+1}$. We use $w_{-1}=0.8$m and choices of $w_1=1.6$m and $3.2$m respectively. Accessible points are left white, grey areas can not be occupied by evacuees.}
\label{fig:fig1}
\end{figure}

In our model we consider situations in which $N$ pedestrians evacuate a rectangular shape with two asymmetric exits, as shown in Fig. \ref{fig:fig1}. Such configurations may be thought of as modelling for example a bridge or a large rectangular corridor or room.
We will refer to the central area of the structure in Fig. \ref{fig:fig1} as the main evacuation area, more precisely this is the rectangle with coordinates $10$m$\leq x\leq 60$m, and $10$m$\leq y \leq 20$m.  Agents in this area will aim to move to the right or left, depending on the choice of exit they have made. Accessible points with $x<10$m are referred to as `Exit route -1' in the following, accessible point with $x>60$m as `Exit route +1'. The choice of notation $\varepsilon=\pm 1$ to denote the two exit routes instead of the more intuitive labels $\varepsilon=1,2$ is made purely to ease the notation in later parts of the paper. Individuals in the areas $x<x_{-1}$ and $x>x_1$ respectively are assumed to have left the central region of evacuation, they aim to move either up or down, as shown in the figure.

In order to propagate the evacuees in space the evacuation area is divided into a homogeneous and isotropic grid of cells. At any one time, each cell of the grid can be either vacant or occupied by a single pedestrian. The size of each cell is chosen to be the typical size that a pedestrian occupies in dense crowds \cite{cell_size}, i.e. roughly $0.4$m$\times 0.4$m. 

\subsection{Update rules of the cellular automaton}
In our simulations agents update their positions in the grid simultaneously, i.e. we choose parallel update dynamics. At any one time each agent has a desired direction of motion (towards one of the two exit routes), defining a most preferable cell to move into, along with a second and third choice of cells. This will be detailed below. A given agent will then either stay in their current cell if all the cells that they are allowed to move to are occupied, or he can move. This then introduces the following notion of velocity: in each time-step a pedestrian can move either by $0.4$m towards the exit or he or she stays put. The velocity of pedestrians under normal conditions has been reported to be between $1-1.5$m/s \cite{HFV00}. We therefore assume that $3$ time steps in our simulation correspond to $1$s, and thus the maximum speed of a pedestrian in our simulations will be $1.2$m/s. 

\begin{figure}
\centerline{\includegraphics[scale=0.325]{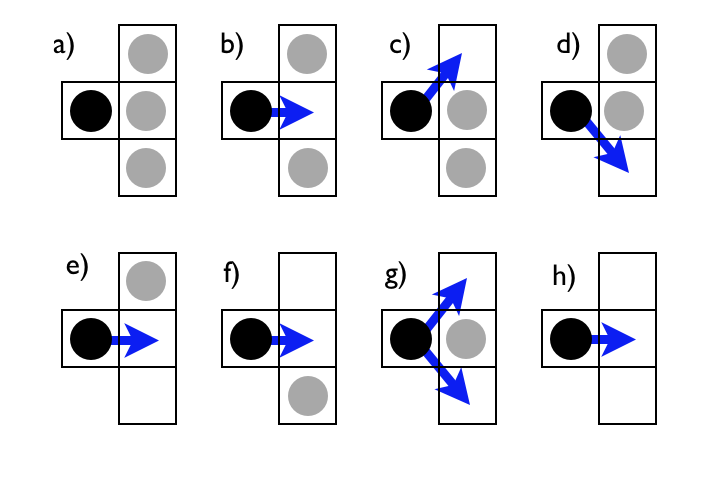}}
\caption{Possible moves in the cellular automaton model. The illustration depicts possible situations a right-moving agent (black disk) can face in the central region of the evacuation scenario. The grey disks depict other agents. The evacuee under consideration can move to one of the three cells on the right whenever these are not occupied. Arrows indicate the moves the agent will make in the different scenarios. In panel a) the agent cannot move. The motion shown in panel b) shows a move towards the agent's preferred cell.  In cases b)-d) only one target cell is vacant. In situations e), f) and h) the preferred cell (horizontal move) is available. In case g) the preferred cell is occupied, but the other two target cells are vacant, so the agent will chose one of the two cells indicated at random (with equal probabilities).}
\label{fig:fig2}
\end{figure}

We will now describe how agents move, starting with evacuees in the central area of the structure. Our model is similar to that of \cite{similar,fukui1,fukui2}. At any point in time each agent in the main evacuation area will have a desired direction of motion, either to the left or to the right. We here describe a right-moving agent, analogous statements apply to a left-moving individual. The states of three cells are taken into account when determining the move of any agent. These are the cell directly ahead of the agent (i.e. to the right in our example), and the two cells above and below this cell ahead, see Fig. \ref{fig:fig2}. The collection of these three potential destination cells can be in one of eight states, each cell either being occupied or vacant. In the situation shown in the figure the agent can move only if he is in any of the states b)-h), in situation a) all possible destination cells are occupied. An agent in the main evacuation area chooses the three possible cells he may move to. When more than one cell is available then the agent prefers to move horizontally. If two cells are equally desirable (see case g)) then he randomly chooses between them. When agents are either in the area of Exit $-1$ or in the area of Exit $1$ then the roles of the vertical and horizontal directions are reversed compared to the picture in the central region. Agents then choose to move upward or downward simply based on closest distance, and the same CA-rules as above apply, with the vertical and horizontal directions interchanged. In the case that more than one agent choose to move an empty cell, one of the candidate agents is randomly selected to move towards the desired empty cell and the others stay in their current position.

\subsection{Decision making and utility}
When pedestrians are either in the area of Exit $-1$ or Exit $1$ they choose to move towards the nearest exit, i.e. up or down. For the pedestrians in the main evacuation area we will introduce a different decision making mechanism, based on the evaluation of a utility function. Agents assign a utility to each exit and they then choose to move to the exit with the maximum utility. For example, the expected time for a pedestrian to reach a given exit could constitute a (simple) utility function. This choice will serve as our starting point. In addition the utilities agents assign to the different exit routes is influenced by potential communication between the pedestrians, i.e. by information they may receive from others. More formally, the utility agent $\alpha$ assigns to  exit $\varepsilon=\pm1$ is given by:
\be
u_{\alpha,\varepsilon}=\frac{r_{\alpha\varepsilon}}{v_{\alpha\varepsilon}},
\label{eq:eq1}
\ee
where $r_{\alpha\varepsilon}$ is the distance of the pedestrian $\alpha$ to exit $\varepsilon$, and where $v_{\alpha\varepsilon}$ is the agent's estimate of the velocity with which they can move to this exit.  
We distinguish two distinct groups of evacuees in the proposed model: (i) Agents without communication:  Agents of in this group choose to move towards their closest exit and they never change this decision.  (ii) Agents with access to communication: The second group consists of agents who have access to means of communication, for example pedestrians who are able to use their cell phones to talk to their peers in the area. Such agents can communicate and exchange information about their velocities and which exit they heading towards. In our model we endow a fraction $c$ of agents with this capability ($c\in[0,1]$). The resulting $cN$ evacuees are then `paired up' at the beginning of the simulation, i.e. each communicating agent is assigned one peer, and is assumed to be communicating with them throughout the evacuation. If, at some point during the evacuation, two communicating agents have chosen different exits as their respective destinations then they can obtain information about the respective other exit route from the communication with their partner. Assume that agent $\alpha$'s direction is towards exit $\varepsilon$ with velocity $v_{\alpha\varepsilon}$, and his peer $\beta$ is heading towards
exit $-\varepsilon$ with velocity $v_{\beta,-\varepsilon}$. Agent $\alpha$ then takes this information into account and evaluates his utilities as follows:

\bd
u_{\alpha,\varepsilon}=\frac{r_{\alpha\varepsilon}}{v_{\alpha\varepsilon}}, ~~~ u_{\alpha,-\varepsilon}=\frac{r_{\alpha,-\varepsilon}}{v_{\beta,-\varepsilon}},
\ed  
agent $\beta$ would assign utilities analogously. The velocities $v_{\alpha,\varepsilon}$ here represent the average forward motion (towards the target) over the last $3$ iteration steps, so it is an estimate of the agent's instantaneous speed.
\begin{figure}
\centerline{\includegraphics[scale=0.5]{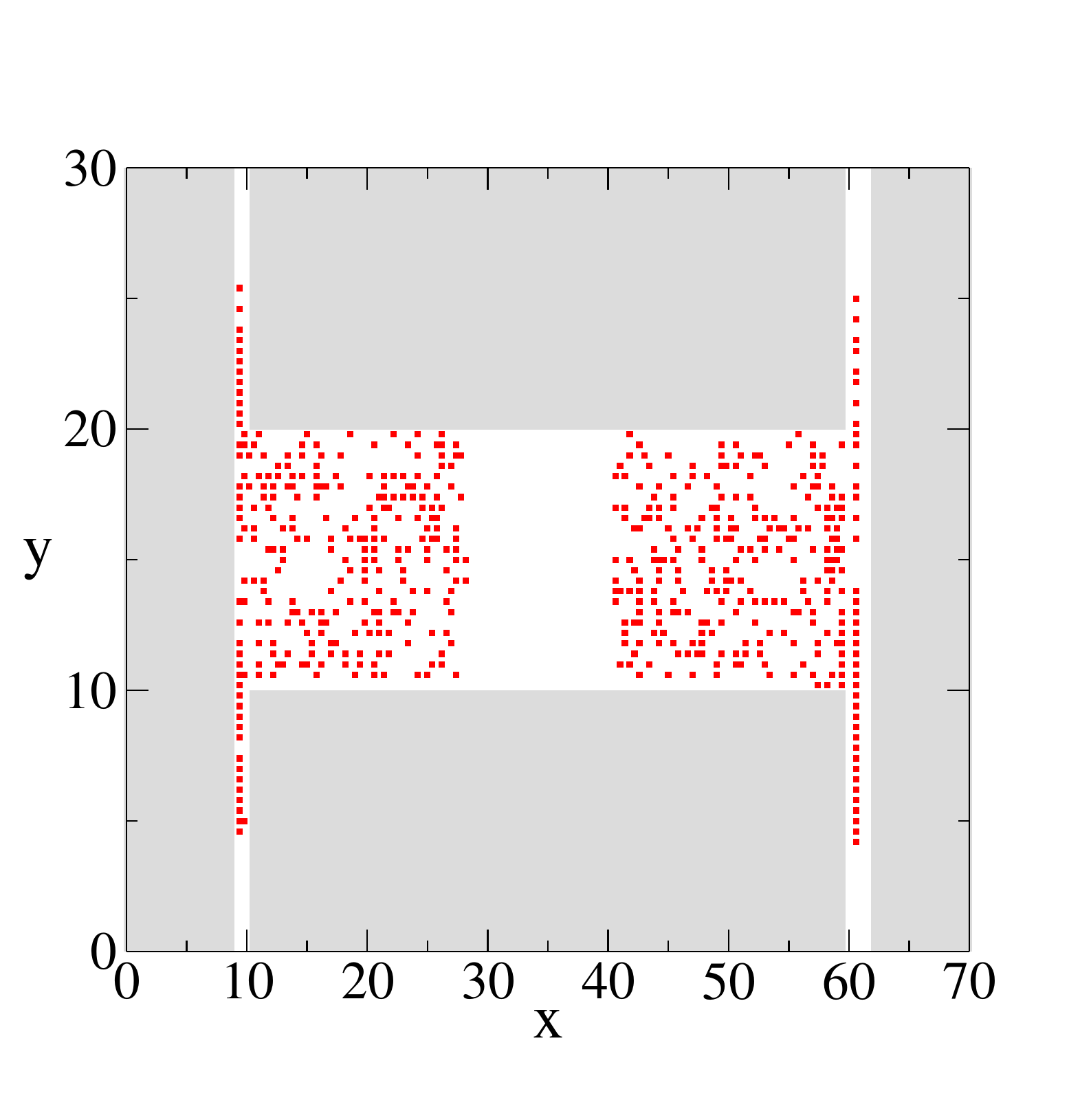}}
\centerline{\includegraphics[scale=0.5]{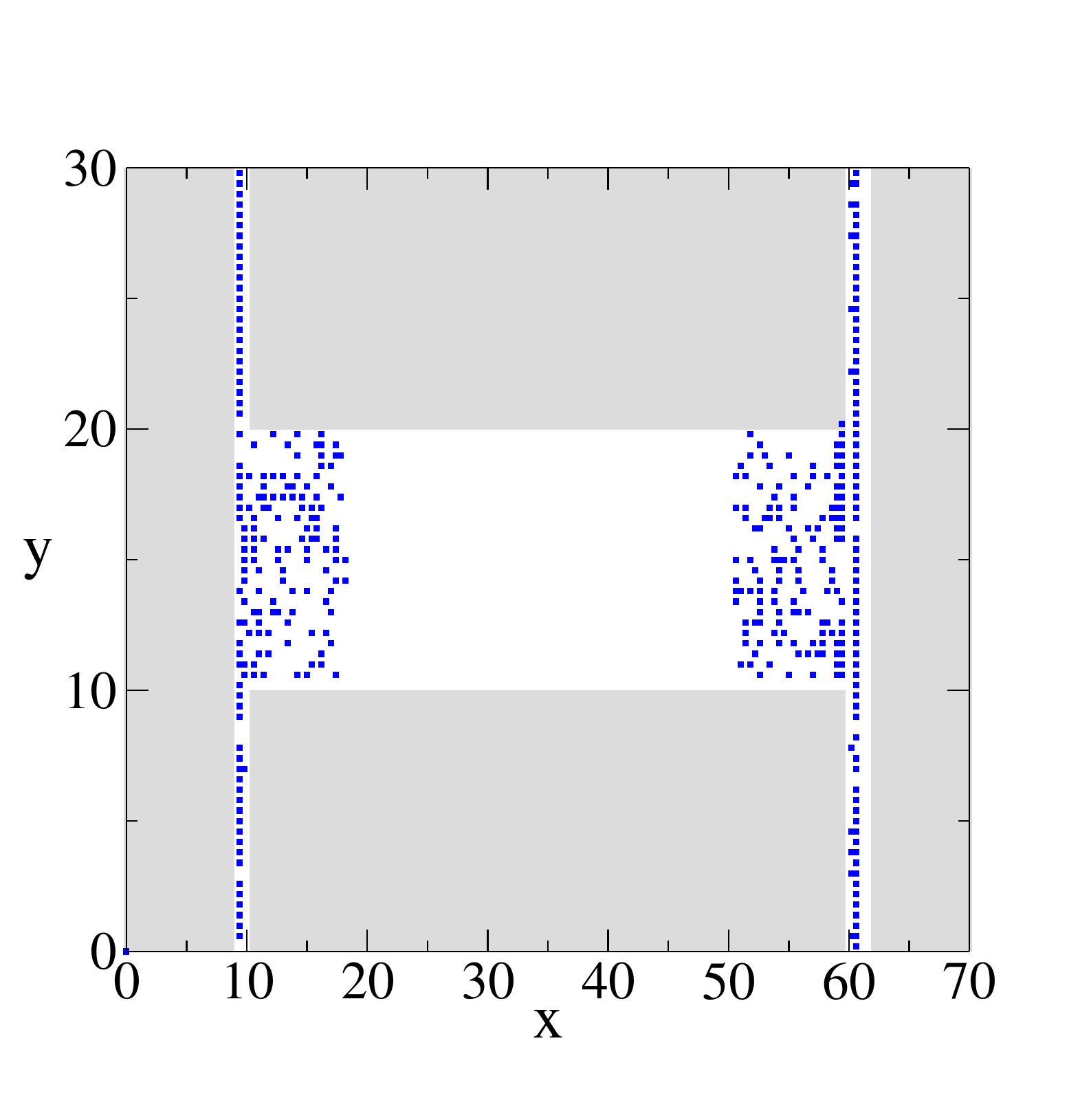}}
\caption{Two snapshots of a simulation with parameters $w_1=1.6$m, $\rho=0.19$ ($600$ evacuees), $c=0.6$, changes of the direction of motion are allowed after $30$s. The snapshot in the upper panel is taken after $15$ iterations of the CA, i.e. after $t=5$s, the one in the lower panel after $40$ iterations ($t\approx13$s).}
\label{fig:fig3}
\end{figure}
In our simulations we do not allow for communication in the first $30$ seconds of the simulation, corresponding to the $90$ initial parallel update steps. Subsequently communicating agents are in permanent contact with their peer. While non-communicating agents pick and exit route at the start and then never change their destination subsequently, agents with the capability of communication may change their preferences as time goes by and in the light of the information they receive. For example their own velocity may become very small as the evacuation proceeds, while the velocity of their peer (if heading to the opposite exit) may remain high. In this case changes of direction may occur. In order to prevent too frequent a change of mind of a given agent we do not allow for further changes in their destination in a certain period of time immediately after a change of direction has been made, details are specified below. We believe that this is a realistic constraint as agents in real-world evacuations will first re-evaluate their velocity, and then only acquire the potential of changing their mind again after some time has elapsed.

\section{Simulation results}\label{sec:res}
\begin{figure}
\centerline{\includegraphics[scale=0.5]{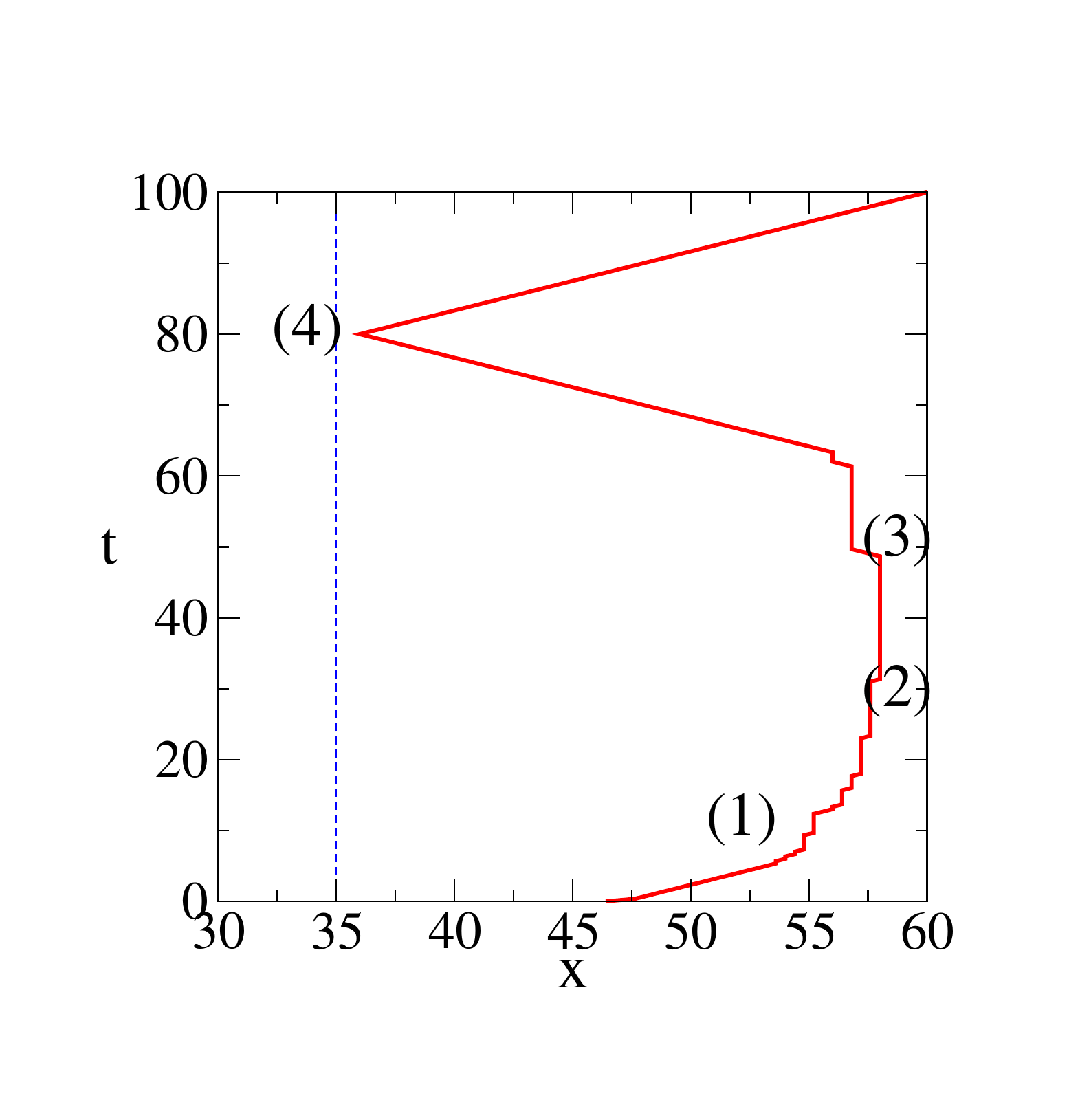}}
\caption{Example trajectory of an agent in a simulation with $N\approx2100$ evacuees ($\rho=0.67$), in a structure with $w_1=1.6m$, and where $60\%$ of agents have access to communication ($c=0.6$). No changes in direction are allowed in the first $30$s of the simulation, and for $30$s after any change in direction has occurred. The vertical dashed line indicates the centre of the structure, both exit routes are equally distant from this line. The trajectory can be divided into four segments, as indicated in the figure: (1) the agent is closer to Exit $1$, and therefore moves to the right; (2) The agent slows down due to jamming; (3) acting upon received information they thus turn back at time $t\approx 50$s, heading to the left; (4) once $30$s have elapsed since the change of direction at (3) the agent still finds themselves closer to Exit $1$ than to Exit $-1$, hence changing direction again.}
\label{fig:fig4}
\end{figure}

In all simulations we use the configuration depicted in Fig. \ref{fig:fig1}. The main evacuation area is divided into a grid of $125\times 25=3125$ cells. The width of the narrow exit route ($\varepsilon=-1$) is chosen to be $w_{-1}=0.8$m. The width of the wide exit, $w_{+1}$ is generally be taken to be $w_1=1.6m$ (unless stated otherwise), but in some simulation scenarios we also varied this parameter. parameters. This is then stated explicitly. As a second control parameter we varied the initial number of evacuees, i.e. the density of occupied cells in the central area at the start of the simulations. This density will be denoted by $\rho\in[0,1]$ in the following, i.e. initially any particular cell in the central evacuation area is randomly populated with probability $\rho$, and initialized in an empty state with probability $1-\rho$. Cells are here treated independently. The simulation then proceeds according the rules specified above, snapshots from a single simulation run are shown in Fig. \ref{fig:fig3}.  In order to illustrate the effects of communication and the decision making of agents we show a sample trajectory of an agent who changes direction multiple times in Fig. \ref{fig:fig4}. As explained in more detail in the figure caption the agent initially moves towards exit $\varepsilon=1$, he is then slowed down due to jamming, and turns back after approximately $50$s of the simulation, based on information received about exit $\varepsilon=-1$ via communication with another agent. After a further $30$s have elapsed the agent is still closer to exit $\varepsilon=1$ than to exit $\varepsilon=-1$, and hence changes direction again at $t\approx 80$s. Note that just before this final change of direction the agent himself and his partner only have information about motion towards Exit $-1$, so that communication does not provide any additional information and decisions are made purely based on distance.

\begin{figure}[t!]
\centerline{\includegraphics[scale=0.5]{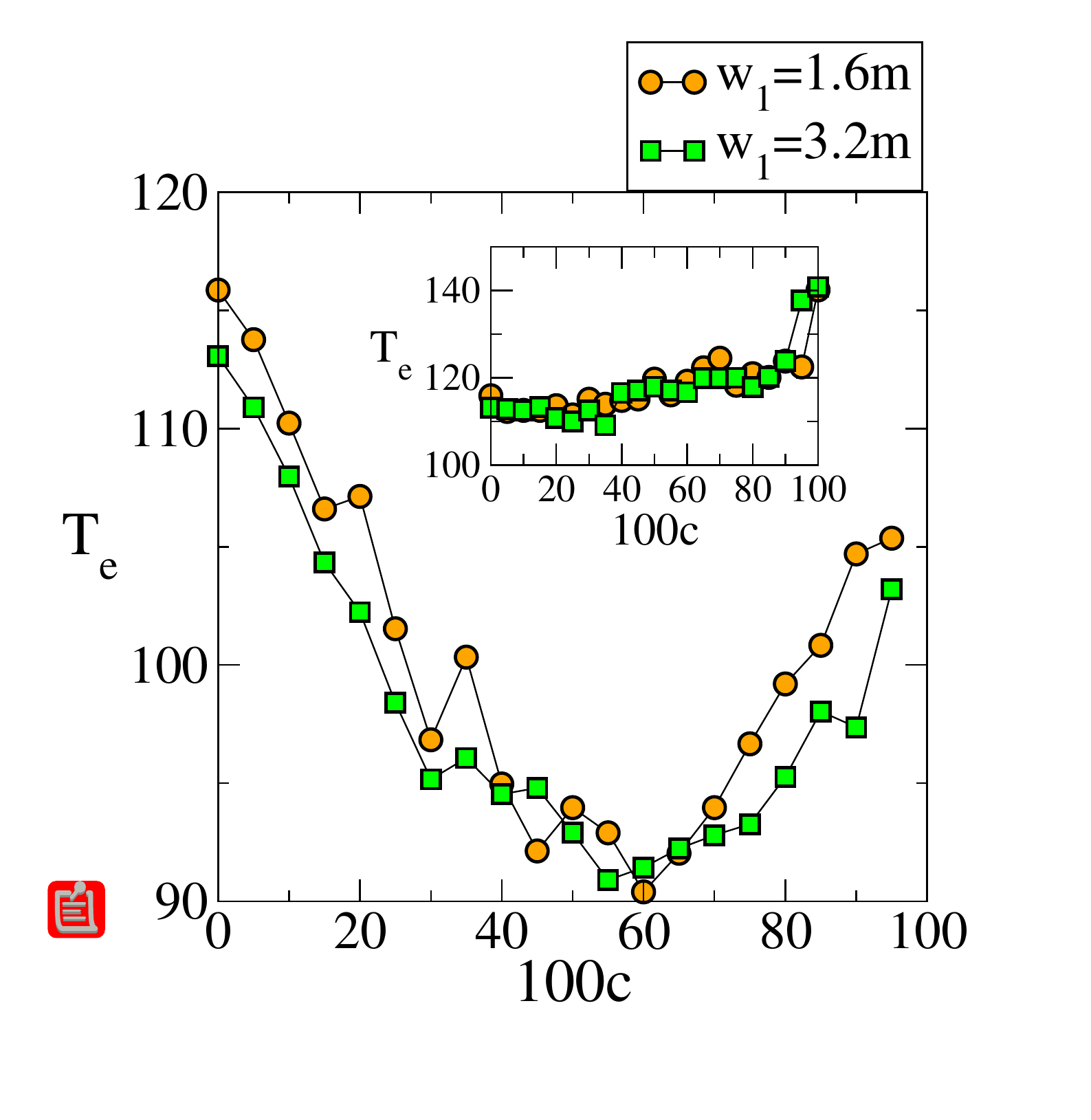}}
\caption{Time needed to evacuate all agents in the structure as a function of the fraction, $100c$, of people who can communicate. Data is an average over $20$ replications of the simulation, and is shown for 2 different configurations, $w_1=1.6$m and $w_1=3.2$m. The initial density of evacuees is $\rho=0.67$. Main panel: changes in direction are allowed once every $90$ iterations of the CA, i.e. once every $30$s. Inset: changes are allowed every $40$ iterations ($\approx 13$s).}
\label{fig:fig5}
\end{figure}

In order to explore the effects of communication and decision making more systematically we have investigated how access to communication affects the total evacuation time. To this end we have varied the fraction, $c$, of the evacuees who are able to communicate, and have measured the time required for the last agent to leave the structure. We perform these experiments using two different configurations, in the first configuration the wide exit has a width of $w_1=1.6$m and in the second configuration we have $w_1=3.2$m. Results for the total evacuation time are shown in Fig. \ref{fig:fig5}, averaged over $20$ independent replications of each simulation trial. We note that the quantity shown in the figure is the time needed to evacuate {\em all} agents, i.e. results represent the time it takes the last agent to leave the structure. When we consider the case where changes in direction are allowed once every $90$ iterations, for both widths of Exit $\varepsilon=1$, we observe that evacuation time is relatively large when either very few or very many agents have access to communication. At intermediate values of $c\approx 0.6$, however, we find a significant reduction in evacuation time. When agents are allowed to change direction more frequently, see the inset of Fig. \ref{fig:fig5}, we observe no minimum at intermediate values of $c$, instead the evacuation time increases steadily as the fraction of communicating agents increases. We note that the bulk of the increase occurs at large values of $c$, whereas the curve is essentially flat at lower fractions of communicators.

\begin{figure}[t!!!!!!!!!!!!]
\centerline{\includegraphics[scale=0.5]{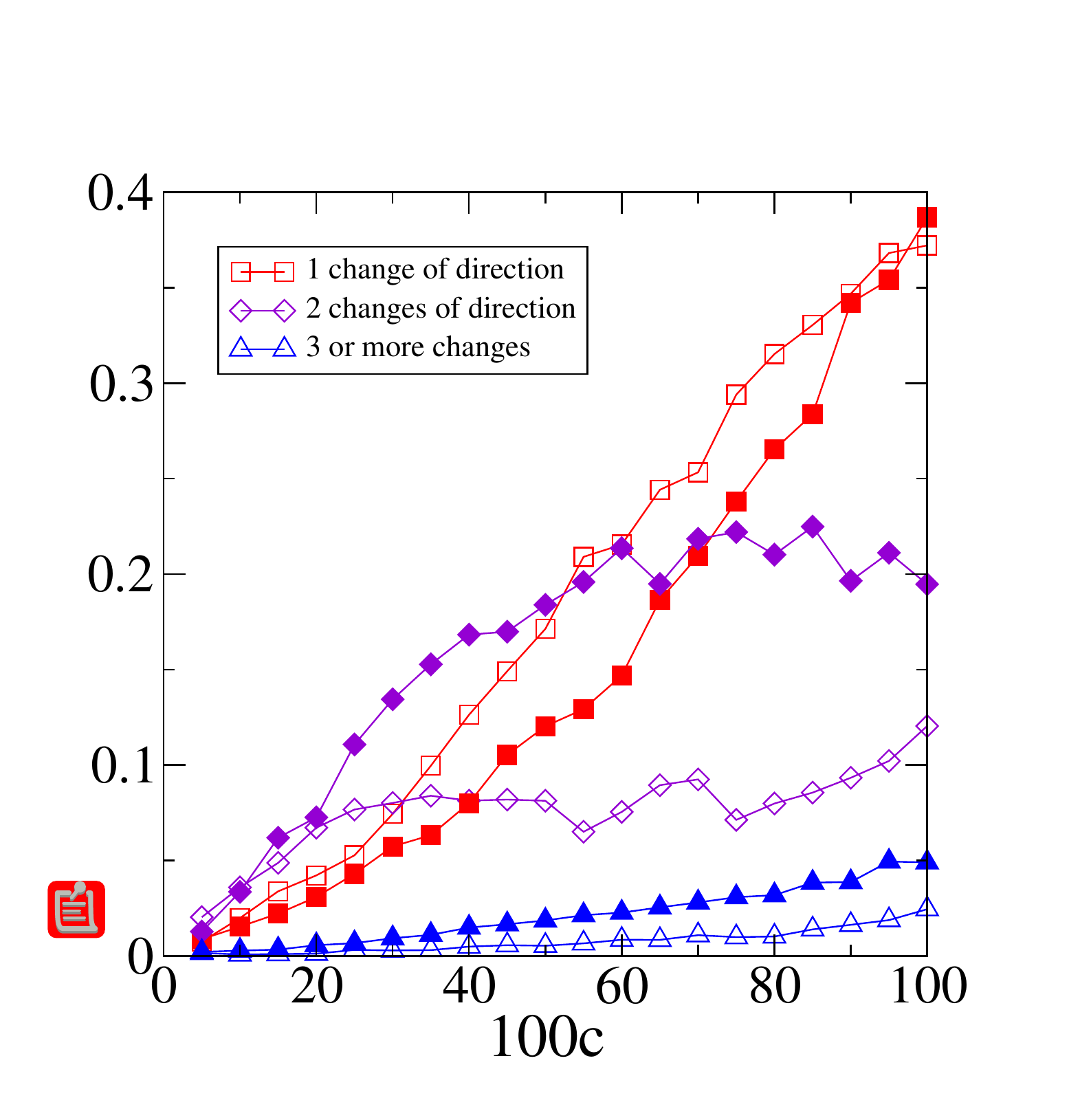}}
\caption{Fraction of agents who change their direction of motion once, twice and three or more times respectively. Simulations are for a density of $\rho=0.67$ ($2100$ agents). Open symbols represent the case in which communication and changes of direction may occur once every $30$ seconds ($90$ iterations), filled symbols correspond to the case in which changes in direction may occur once every $13$ seconds ($40$ iterations). }
\label{fig:figg6}
\end{figure}

\begin{figure}[t!!!!!!!!!!!!]
\centerline{\includegraphics[scale=0.45]{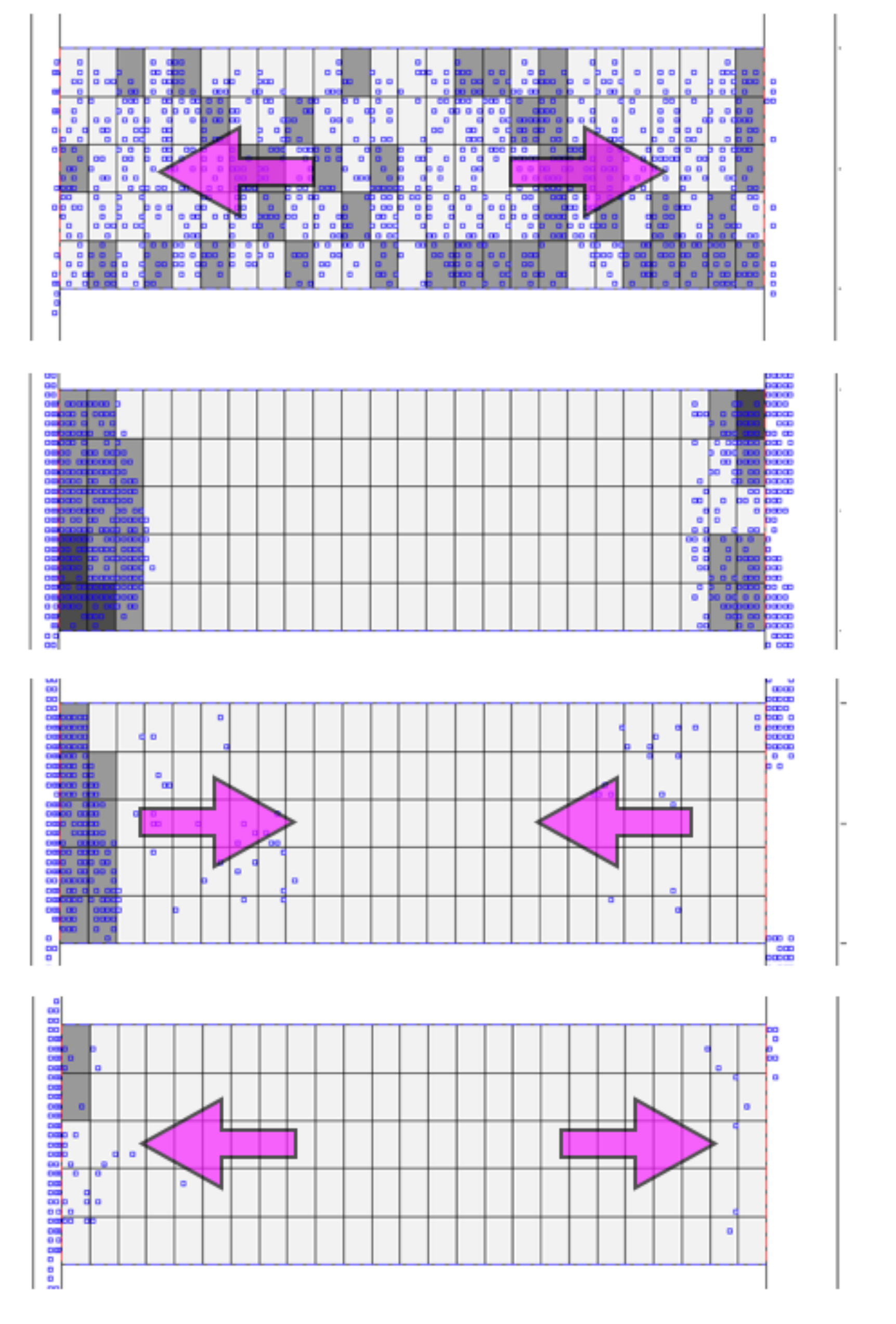}}
\caption{Dynamics of a simulation with $2100$ agents, $60$ percent of which have access to communication. The four snapshots are taken at the beginning of the simulation, then after $50$ iterations of the cellular automaton (corresponding to approximately $16$ seconds), and further after $75$ and $120$ iterations respectively ($25$ and $40$ seconds). For reasons of clarity only a small fraction of all agents ($\approx 20\%$) is displayed. The shaded cells in the background indicate the local density of agents (on a coarse-grained scale), with darker colours indicating higher densities. The arrows show the dominant direction of motion.}
\label{fig:figg7}
\end{figure}

The observations of Fig. \ref{fig:fig5} may be interpreted as follows. The initial decrease of evacuation time in the data shown in the main panel is due to the effects of communication. In the initial parts of the simulation all agents head towards the nearest exit, so approximately half of the agents will move towards the narrow exit, which then becomes jammed. As the number of communicating agents increases more and more of those agents revise their decision, and move towards the wider, and less jammed, exit. This is also confirmed in Fig. \ref{fig:figg6}, in which we show the fractions of agents who change their direction once, twice and three or more times. When the proportion of communicators is too large an increasing fraction of agents makes multiple changes of their direction (see Fig. \ref{fig:figg6}), which ultimately leads to a slowdown, and hence an increase in evacuation time. This, we believe, is the reason for the increase in $T_e$ shown in the main panel of Fig. \ref{fig:fig5} above $c\approx 0.6$. The behaviour of agents is also illustrated in Fig. \ref{fig:figg7}. For reasons of clarity we only show a small fraction of all agents, but as seen in the figure the ability to communicate does not only lead to multiple changes of direction, but also to a counterflow in the centre of the structure. This can lead to additional jamming and further changes in direction, adding to the slowdown of the dynamics. 

In the inset of Fig. \ref{fig:fig5}, corresponding to a simulation in which agents are allowed to exchange information more frequently than in the simulations shown in the main panel, no minimum in evacuation time is observed as a function of $c$. In this scenario evacuees intrinsically have a higher propensity to change directions (due to the frequency with which they may revise their decisions) and hence the slowdown effect sets in at a lower value of $c$ than in the main figure, overriding the speed-up effect. This is confirmed again in Fig. \ref{fig:figg6}, the data shows that the fraction of agents changing directions multiple times is increased when information exchange is allowed once every $40$ iterations relative to the case in which communication is allowed once every $90$ iterations.


\begin{figure}[t!!!]
\centerline{\includegraphics[scale=0.5]{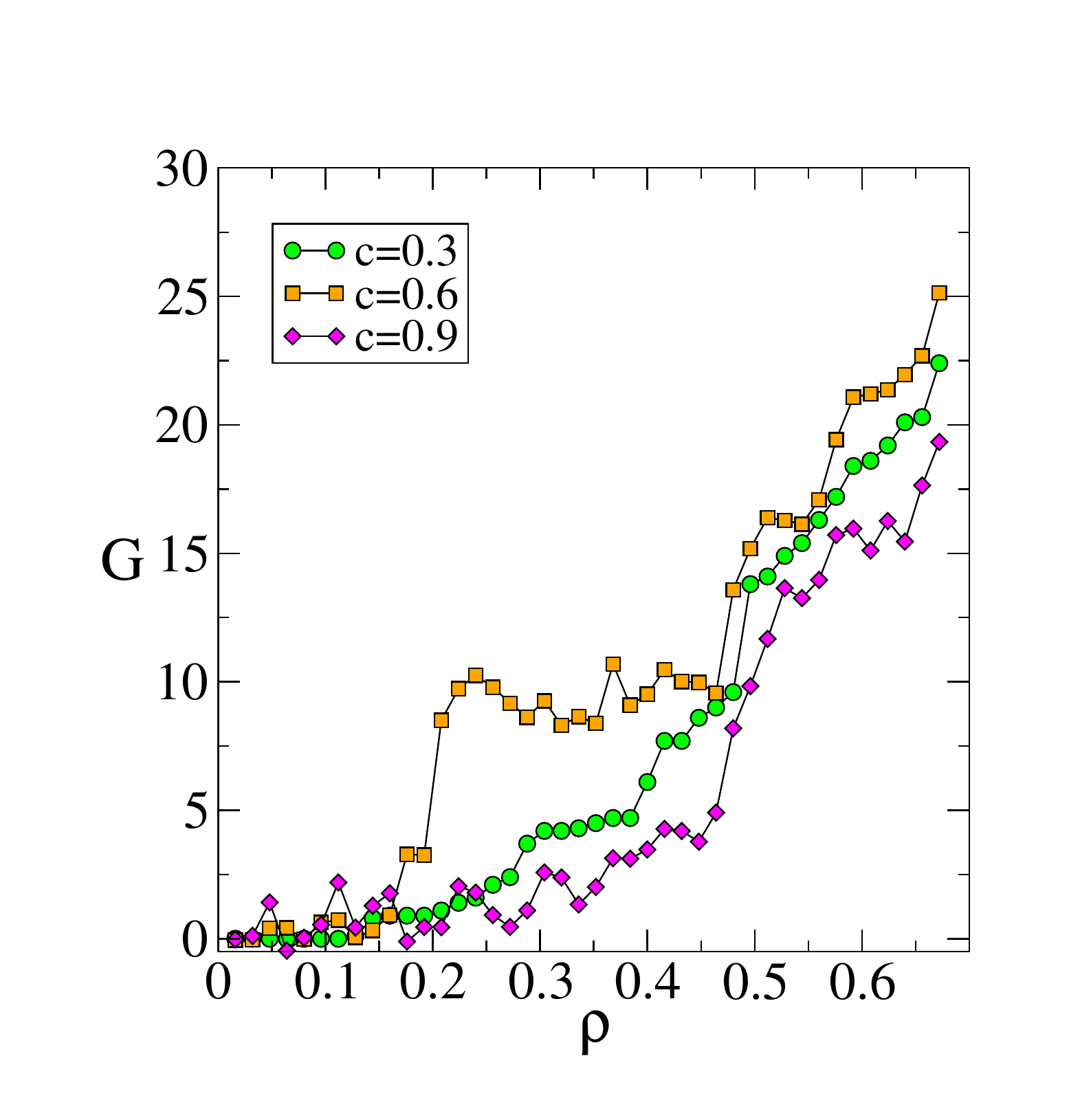}}
\caption{Relative gain in evacuation time (in percent) when comparing situations in which $30\%$, $60\%$ and $90\%$ of agents, respectively, communicate against a baseline of no communication. Data is shown as a function of the initial density, $\rho$, of agents. Markers are an average over $100$ replications of the simulation. Changes in direction are allowed once every $30$s.}
\label{fig:figg8}
\end{figure}

In order to obtain a more detailed dynamical picture of the evacuation dynamics in our model, we have also studied the effects of the initial crowd density in the main evacuation area, $\rho$, on the performance of the overall system. For different densities we have here compared the performance of the evacuation procedure using three fractions of communicating agents, $c=0.3, 0.6$ and $c=0.9$. We denote the corresponding evacuation times by $T(c)$, and study the gain $G(c)=100\times \frac{T(0)-T(c)}{T(0)}$, relative to a baseline of no communication at all ($c=0$). This is then an indicator of how much communication can speed up the evacuation dynamics, the quantity $G(c)$ represents the (relative) gain in evacuation time in percent. Results are shown in Fig. \ref{fig:figg8}, the reported data are averages of $100$ independent simulation runs, to reduce statistical errors. As seen in the figure communication does not appear to have any noticable effect at low densities of pedestrians, $\rho\lesssim 0.2$, but it can lead to significant reduction in evacuation time at higher densities. Intuitively such behaviour is not unexpected, at low densities agents essentially move freely in the structure, and hence there is no reason for them to change direction during the process of the evacuation. At an intermediate density $\rho\approx 0.2$ we can observe a jump in gain in Fig. 8 (see the curve for $c=0.6$). This discontinuity is a reflection of the jamming occurring at the narrow Exit ($\varepsilon=-1$) beyond a certain critical density. Agents heading towards Exit $-1$ will be able to move only with small velocity because of congestion and hence those who have access to communication choose to change direction, and so the gain due to communication is enhanced. At high densities $\rho\approx 0.5$ we observe a further increment of $G$, this is when the wider exit becomes jammed as well, so an additional fraction of agents will be able to benefit from the ability to communicate and change their direction if needed. At the highest density we have looked at ($\rho=0.67$) we find a total
gain of up to 25\% in our model, the gain is maximised at approximately $c\approx 0.6$, as also seen in Fig. \ref{fig:fig5}.

 \begin{figure}[t!!!!!!!!!!!!]
\centerline{\includegraphics[scale=0.5]{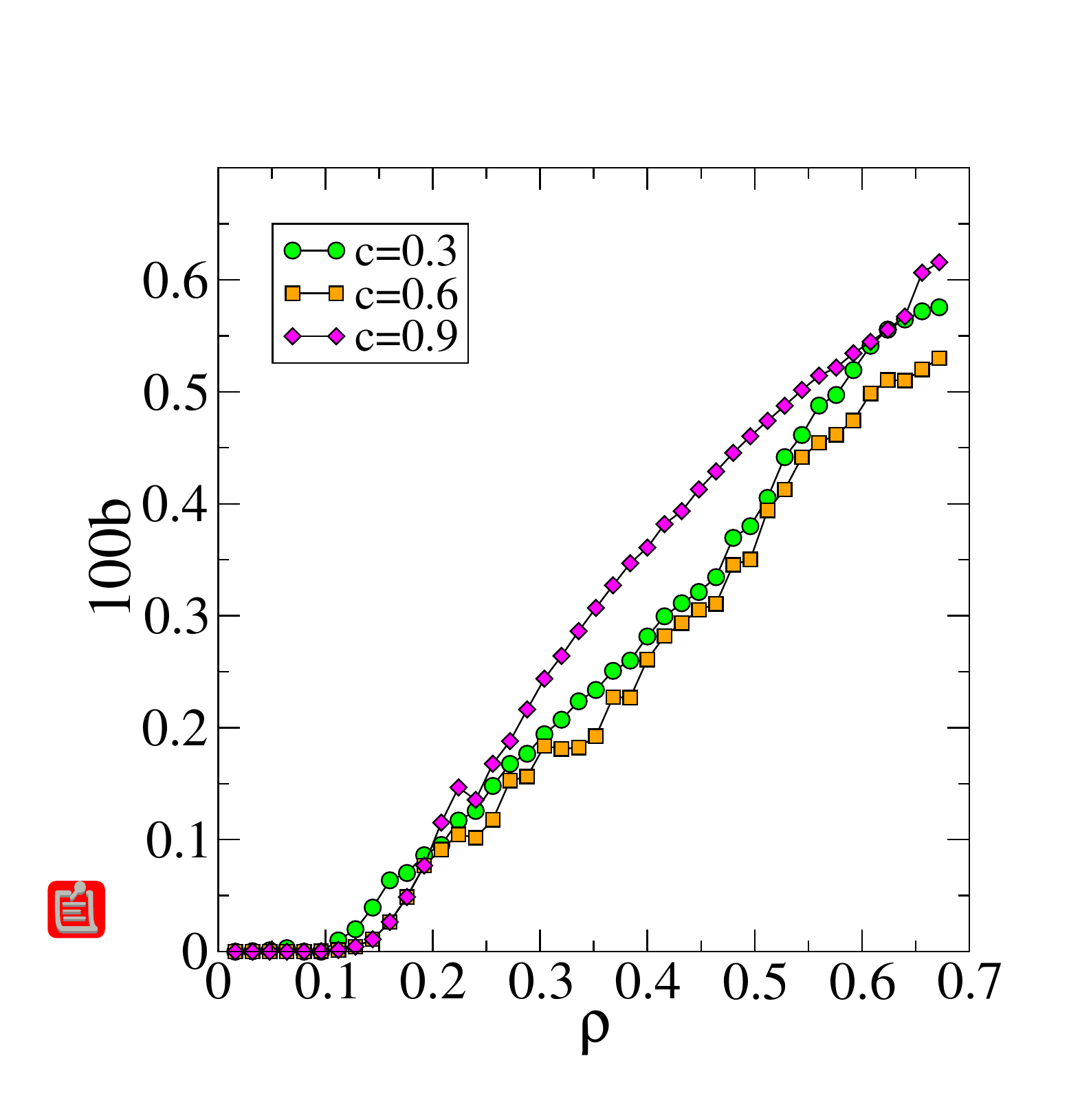}}
\caption{Time-averaged percentage of non-movers, $100b$, as a function of the initial density, $\rho$, of agents, when the percentage of communicating agents is $c= 30, 60$ and $90 \%$ respectiely. Markers are an average over $100$ replications of the simulation. Changes in direction are allowed once every $30$s.}
\label{fig:figg9}
\end{figure}
In Fig. \ref{fig:figg9} finally we show the fraction of number of jammed agents. At each time step of the simulations we measure the number of pedestrians $N_b(t)$ who can not move, and hence obtain $b(t)=N_b(t)/N(t)$, where $N(t)$ is the number of evacuees still within the structure at this time (agents who have already evacuated are disregarded). The data in Fig. \ref{fig:figg9} then represents the time average $b=\avg{b(t)}$ over the entire simulation when the percentage of communicating agents is $c= 0.3, 0.6$ and $0.9$ respectiely.  The data confirms the jamming transition at $\rho\approx 0.2$, and also illustrates again that an intermediate fraction of communicators minimizes blockages, the fraction of non-movers $b$ is consistently lower for $c=0.6$ than for $c=0.3$ and $c=0.9$.      

\section{Conclusions and outlook}\label{sec:concl}
In summary we have used a simple cellular automaton model of pedestrian motion to study the effects of decision making and communication on the outcome of evacuation from a spatial structure with two exit routes. Initially agents move towards the nearest exit in our model, but they may revise their decision at later times, based on utility functions they carry for each of the two exits. These utilities take into account the distance from the respective exits, and the perceived velocity with which any particular agent may move towards the exit routes. We here adopt a basic decision rule, in which an agent simply moves towards the escape route with the higher (perceived) utility. Extensions to stochastic decision making, for example based on logit rules are possible \cite{logit}. As one main ingredient of our model agents can communicate with other evacuees as they move through the structure. Specifically we have communication via mobile phones in mind, but the concept is generic and can be extended to other channels, e.g. text messaging or social networks. Individuals with access to communication take into account the information they receive when evaluating the utilities of the two exit routes. With time their own progress in moving towards their preferred exit route may develop either for the better (the agent moves with a reasonably high velocity) or for the worse (the agent is stuck). Similarly the information they receive may change as the simulation progresses, for example they may receive word that the flow at the respective other exit is relatively free from congestion. Based on these factors agents may revise their decisions and reverse their preferred direction of motion. In order to prevent unrealistically frequent changes of mind, we have introduced a minimum time lag between such events, typically $30$s or so. 

As a key result of our simulation study we find that access to communication can significantly reduce the total time needed to evacuate all agents from the structure. Interestingly one finds that evacuation proceeds the quickest at intermediate numbers of communicating agents, typically $60\%$ or so in the cases we have tested. If only a small number of individuals has access to communication agents essentially choose the nearest exit at the beginning, and are then unlikely to ever turn back. This can lead to jamming. If a very high proportion of agents can communicate, however, then a large fraction of agents has information about both exit routes, and hence there is a significant probability that large groups of agents make similar decisions, again leading to jamming at the exit chosen by these groups. Agents then also have a propensity to change direction multiple times, leading to loss of time and to counterflows detrimental to the overall outcome. At intermediate access to communication both effects are absent (or minimized), leading to a relatively smooth evacuation flow. The observed reduction of evacuation time due to communication is however crucially dependent on the initial number of individuals present in the area to be evacuated. If this area is only sparsely populated from the start, then jamming does not occur in the first place, so that even in absence of communication the evacuation proceeds smoothly. Interestingly we find two critical values for the density of pedestrians, separating dynamical regimes in which none, one or both exit routes become at least partially jammed.
It is important to stress though that the underlying cellular automaton used in our work to propagate particles in space is a relatively simple one, and hence presumably not very realistic in comparison to more involved models. In particular, because of the configuration we selected, we have not attempted to implement approaches based for example on floor fields \cite{floor}. The main contribution of our work is not to perfect what is referred to as the `operational level' of evacuation simulation \cite{schadschneider}, i.e. the mechanics by which individuals move. Instead we aim to address decision making (referred to as the `tactical level' \cite{schadschneider}), and in particular real-time communication during the evacuation event. Even in the basic model we have considered here communication can have a significant effect on the overall time needed to evacuate simple structures. Given these promising results we believe that the utility-based approach put forward here can make useful contributions to modelling these aspects. We therefore believe that our work can reasonably be expected to serve as a base for including such elements in evacuation models with more intricate operational-level dynamics.

%%%%%%%%%%%%%%%%%%%%%%%%%%%%%%%%%%%%%%%%%%%%%%%%%%%%%%%%%%%%%%%%%%

\begin{acknowledgments} 
TG would like to thank Research Councils UK for support (RCUK reference EP/E500048/1). The authors acknowledge funding by the Engineering and Physical Sciences Research Council (EPSRC), under grant EP/I005765/1 (IDEAS Factory - Game theory and adaptive networks for smart evacuations). We would like to thank D. Helbing, I. Farkas, N. Jones, M. Kolokitha and J. Preston for discussions.

\end{acknowledgments}

\end{document}